\documentclass[aps,pra,preprint,showpacs]{revtex4}
\usepackage{amsmath}
\numberwithin{equation}{section}
\usepackage{graphicx}
\usepackage{comment}
\usepackage{url}

\begin{document}
\title{A comparison between detailed and 
configuration-averaged collisional-radiative codes 
applied to non-local thermal equilibrium plasmas}

\author{M. Poirier}
\author{F. de Gaufridy de Dortan}
\affiliation{
Commissariat \`a l'\'Energie Atomique, 
Service ``Photons, Atomes et Mol\'ecules'', 
Centre d'\'Etudes de Saclay, 
F91191 Gif-sur-Yvette \textsc{Cedex} France }
\date{\today}
\begin{abstract}
A collisional-radiative model describing non-local-thermodynamic-%
equilibrium plasmas is developed. It is based on the HULLAC suite 
of codes for the transitions rates, in the zero-temperature 
radiation field hypothesis. Two variants of the model are 
presented, the first one is configuration-averaged, while the 
second one is a detailed level version. Comparisons are made 
between them in the case of a carbon plasma; they show that the 
configuration-averaged code gives correct results for an 
electronic temperature $T_e =\text{10 eV (or higher)}$ but fails 
at lower temperatures such as $T_e =\text{1 eV}$. The validity of 
the configuration-average approximation is discussed: the intuitive  
criterion requiring that the average configuration-energy 
dispersion must be less than the electron thermal energy turns out 
to be a necessary but far from sufficient condition. Another 
condition based on the resolution of a modified rate-equation 
system is proposed. Its efficiency is emphasized in the case of 
low-temperature plasmas. Finally, it is shown that near-threshold 
autoionization cascade processes may induce a severe failure of the 
configuration-average formalism.
\end{abstract}
\maketitle


\section{Introduction}
It is now well-known that in highly-charged hot plasmas, emission 
and absorption spectra usually display broad structures 
theoretically described as unresolved transition arrays (UTA) 
\cite{Bau79,Bau82,Bau88}, spin-orbit split arrays (SOSA) 
\cite{Bau85} or supertransition arrays (STA) \cite{Bar89}. Isolated 
lines may also be present when transitions occur between two 
configurations of small degeneracy. The UTA formalism consists in 
expanding the individual transition energies as a function of the 
various moments $<E^n>$ --- where the ponderation is performed 
using line strengths --- with the assumption that levels inside a 
given configuration are distributed according to thermal 
equilibrium, the validity of this assumption lying on the condition 
that the configuration width $\Delta_c$ must be smaller than the 
thermal energy $k_BT_e$. The various configuration populations are 
then calculated using averaged rate equations: the collisional 
and radiative transition rates are averaged over the states of the 
initial configuration and summed over the states of the final 
configuration. Another interesting possibility is the description 
of plasmas out of local thermodynamic equilibrium (LTE) in terms 
of \textit{effective} temperatures \cite{Bus93}. 
It has been demonstrated \cite{Bau00} that, in certain 
circumstances, a temperature different from the electronic 
temperature may be defined for each configuration. Such an 
ensemble of non-LTE parameters may be obtained via the resolution 
of an inhomogeneous system of linear equations. However, this 
theoretical derivation emphasizes that the validity of such 
an assumption is restricted to certain types of configurations.
An even more drastic assumption can be made (e.g., \cite{Poi06}): 
one may describe the population of each \textit{ion} according to a 
simple collisional-radiative (CR) model given by Colombant and Tonon 
\cite{Col73}. This simple model includes collisional ionization, 
three-body recombination and radiative recombination, which are the 
only processes possible between the ground states of two ions of 
consecutive charges in the absence of any external radiation field; 
inside each ion, the excited-level populations obey a simple 
Boltzmann distribution. 

Since the early proposals of Bates \textit{et al} \cite{Bat62} 
or McWhirter \cite{MWh78}, numerous collisional-radiative codes 
have been developed. They are based on various atomic descriptions, 
such as hydrogenic approximation for SCAALP (Self-Consistent 
Average Atom for Laboratory Plasmas) \cite{Fau01}, 
supraconfiguration for SCROLL (Super Configuration Radiative 
cOLLisional) \cite{Bar97,Bar00}, AVERRO\`ES (AVERage Rates for Out 
of Equilibrium Spectroscopy) \cite{Pey01}, or MOST (MOdel for 
computing Superconfiguration Temperatures) \cite{Bau04,Han06}, 
parametric or analytical potentials for ATOM3R-OP (code by 
Minguez \textit{et al}, including atomic and optical properties) \cite{Rod06}. Hansen \textit{et al} \cite{Han06} have analyzed 
the respective efficiency of detailed and superconfiguration 
averaged CR models. 

Due to its numerous applications, the theory of non-LTE plasmas is 
nowadays a very active subject \cite{Bow03,Bow06}. For instance, 
laser-produced and discharged-produced plasmas appear as very 
promising sources of intense extreme-UV light well suited for 
13.5 nm-lithography (\cite{Nis06,OSu06,ARa06} and other references 
in the same volume); for such plasmas of moderate density, the 
non-LTE condition prevail in most cases. Another domain of 
application is low-density astrophysics plasmas as well as 
laboratory coronal plasmas.

The aim of this paper is first to present a detailed CR model. One 
essential feature is that it must be based on a \textit{reliable 
atomic code}. To this respect, the HULLAC (Hebrew University 
Lawrence Livermore Atomic Code) parametric-potential code is known 
to be efficient in applications dealing with ion spectroscopy and 
collisional rate calculation. A known limitation of several atomic 
models used in plasma physics \cite{Fau01,Bar97,Bar00,Bau04,Han06} 
is that configuration interaction is ignored. However it has been 
demonstrated that configuration interaction may play a major role 
in several radiative effects in plasmas, especially when 
$\Delta n=0$ transitions are involved, such as in the xenon 13.5 nm 
emission \cite{Gil03,ARa06}. Let us mention that ``configuration 
interaction'' means here interaction in its broader sense, such as 
between $4p^5 4d^9$, $4p^6 4d^7 4f$ and $4p^6 4d^7 5p$ in xenon 
\textsc{xi}, and not what is sometimes called identically but is 
only a restricted interaction between relativistic configurations, 
i.e., a change from pure $jj$-coupling scheme to some intermediate 
coupling \cite{Bau91}. Therefore, it is desirable to build a CR 
formalism based on an atomic model that includes mutual influence 
of configurations. Moreover, it is important to define a correct 
validity condition of such formalisms: in the case of configuration 
average, is it enough to check that the rms deviation of the level 
energy inside a given configuration, ponderated by the 
corresponding configuration population, is less than the thermal 
electron energy? 

The present paper is organized as follows. In Section~%
\ref{sec:codesc}, we review the list of processes accounted for in 
the HULLAC suite and included in the present CR code, with emphasis 
on hypotheses particular to the present formalism. We briefly 
describe the configuration-averaged equations and define a modified 
CR system. Examples of results are then given in 
Section~\ref{sec:carbon} in the case of a carbon plasma, using both 
detailed code and configuration-averaged code. The influence of the 
electron temperature $T_e$ on the configuration-average results is 
discussed in Section~\ref{sec:discus}, pointing out possible 
limitations of this averaging procedure. Conclusions 
and perspectives are exposed in the last section.

\section{
Description of the HULLAC-based collisional-radiative code
\label{sec:codesc}}

\subsection{The HULLAC code}\label{sec:HULLAC}
The HULLAC suite of codes \cite{Bar01} is widely used to study 
ionic spectroscopy as well as collisional processes. Its key 
features among which the fully relativistic formalism, the account 
for interaction between configurations, the efficient description 
of continuum wavefunctions using the phase-amplitude method, the 
fast computation of collisional cross-sections using the 
factorization method, make it a valuable tool in the atomic physics 
of plasmas. Possible alternatives are the codes based on the 
multi-configuration Hartree-Fock (MCHF) or Dirac-Fock (MCDF) 
formalisms, such as the MCHF Cowan's code \cite{Cow81}, or other 
parametric-potential codes such as M. F. Gu's Flexible Atomic Code 
\cite{Gu01,Gu03}, as used in \cite{Rod06}.

We briefly describe below the radiative and collisional processes 
for which the rates are calculated using the HULLAC code. Emphasis 
is put on the assumptions particular to the present work and on the 
implications of the detailed balance principle.

\subsubsection{Radiative bound-bound transitions}
Radiative transition probabilities $g_jA_{ji}$, i.e., Einstein $A$ 
coefficient from the state $j$ to $i$ multiplied by the degeneracy 
$g_j$ of the upper state, are computed for each relativistic level. 
Since in the present formalism we assume that no outer 
electromagnetic field is present, we do not account for absorption 
and stimulated emission rates. This amounts to consider a 
zero-temperature radiation field or optically thin plasmas. 
However the present formalism can deal with non-zero-temperature 
radiation fields too. 

\subsubsection{Autoionization and dielectronic recombination}
The HULLAC code provides the autoionization rates $I_{ij}$ from 
level $i$ to $j$. The rate for the inverse process, dielectronic 
recombination, follows from $I_{ij}$ using the \textit{detailed 
balance principle}
\begin{equation}
	D_{ji} = I_{ij} N_i^\text{SB}/N_j^\text{SB}\label{eqn:aidr}
\end{equation}
where $N_i^\text{SB}/N_j^\text{SB}$ is the ratio of the $i$ to 
$j$ populations \textit{as calculated with the Saha-Boltzmann 
equation for the temperature $T_e$}, but not necessarily equal to 
the observed population ratio,
\begin{equation}
	N_i^\text{SB}/N_j^\text{SB} = \frac{g_i}{g_j} \frac{n_e}{2} 
	 \left(\frac{h^2}{2\pi m k_B T_e}\right)^{3/2} 
	 \exp\left(-\frac{E_i-E_j}{k_B T_e}\right)\label{eqn:sahaboltz}
\end{equation}
where $g_k$ is the $k$-level degeneracy, $E_k$ its energy  
calculated taking the fully-stripped-ion energy as zero, $n_e$ 
and $T_e$ being the electron density and temperature. Since 
collisions are much more frequent between electrons than ions, 
the electrons have been assumed to obey a Maxwellian distribution 
at $T_e$. In some infrequent cases (about 0.1\%) and only for the 
lowest charge states such as C\textsc{i}), very large rates may 
be obtained: such unphysical numbers are simply cancelled, 
which has a very small effect on the CR system solution.

\subsubsection{Collisional excitation and deexcitation}
The collisional excitation cross section is given in HULLAC by a 
four-term fit, according to the Sampson \textit{et al} expression 
\cite{Goe80,Sam83,Fon93}. A drawback of this fit is that it 
sometimes gives rise to negative cross-sections at threshold or 
at large electron energies, and even to negative excitation rates. 
This behavior is connected to the used Sampson fit formula 
\cite{Goe80,Sam83} which is unable to describe forbidden 
transitions or strong configuration mixing: the cross-sections 
may then exhibit a complex energy dependence with multiple maxima, 
not properly described here.

A possible workaround would be to express such cross-sections 
as function of the corresponding radiative rates $A_{ji}$ provided 
the transition is allowed, using, e.g., the van Regemorter 
formulae \cite{vRe62,Mew72}. Another possibility is to use 
alternative fitting expressions as, e.g., splines discussed by 
Burgess and Tully \cite{Bur92}. 

Nevertheless, since this event is only marginal (less than 5\% 
of the cross-sections in cases considered here) these pathological 
transition rates have simply been canceled. Another option would 
be to get a more efficient version of the cross-section fit. 

As above, the inverse process, collisional deexcitation, is given 
by the detailed balance principle
\begin{equation}
	R^{\text{cd}}_{ji} = R^{\text{ce}}_{ij}\frac{g_i}{g_j}
	\exp\left(\frac{E_j-E_i}{k_B T_e}\right).
\end{equation}

\subsubsection{Collisional ionization and three-body recombination}
Collisional ionization is given by a Sampson-type interpolation 
formula \cite{Fon93}. The inverse process, three-body 
recombination, also obeys the detailed balance equation. Similarly 
to the autoionization-dielectronic recombination relation 
(\ref{eqn:aidr}), one gets the three-body recombination rate from 
the collisional ionization rate
\begin{equation}
	R_{ji}^{\text{3br}} = R_{ij}^{\text{ci}} 
	 N_i^\text{SB}/N_j^\text{SB}.
\end{equation}
with the same Saha-Boltzmann population ratio (\ref{eqn:sahaboltz}).
Contrary to collisional excitation, the computed collisional 
ionization rates are always non-negative and do not require a 
special test procedure. The only difficulty may arise from 
transitions at very small energy for which it is unclear whether 
the $j$ state of the next ion is energetically above the $i$ 
state. Let us also mention that, if the $i \rightarrow j$ 
autoionization transition is allowed, the corresponding rates for 
collisional ionization and photoionization are not calculated by 
HULLAC.

\subsubsection{Photoionization and radiative recombination}
Photoionization (PI) cross-sections are obtained from HULLAC 
using a three-parameter formula. One will note 
that in the absence of an external electromagnetic field, 
the PI \textit{rate} is zero. Free-free transitions (inverse or 
direct bremsstrahlung) are not accounted for here. They may be 
included using, e.g., semi-classical description, and would 
give rise to a smoothly varying probability. Besides, since 
here we consider mainly bound level dynamics, they should not 
influence much these populations.

The rate for the opposite process, radiative recombination, is 
again given by the detailed balance equation, 
where the spectral intensity of the outer field is taken equal to 
zero here since no external electromagnetic field is considered. 
Nevertheless, the PI effect will of course be included if one has 
to compute opacities. Here again, due to some limitations 
in the used fit, in some cases the HULLAC rate is singular; 
however, as mentioned below in the case of carbon, such situation 
is even less common than the occurrence of computed negative 
excitation rates described above. The affected recombination rates 
are once again cancelled.

\subsection{The collisional-radiative code}
\subsubsection{Detailed and configuration-averaged CR equations}
Once the various transition rates are computed, one may write 
down the system of rate equations
\begin{equation}
	\frac{dN_i}{dt} = \sum_j R_{ji} N_j - N_i \sum_j R_{ij}
	\label{eqn:dlrate}
\end{equation}
where $R_{ji}$ is the sum of all the rates described above.
Two versions of the code have been developed. This first one 
addresses the direct solution of the system (\ref{eqn:dlrate}), 
which may involve thousands of levels. The second performs 
the configuration average (CA) detailed in Appendix 
\ref{sec:cfavcr}. 
One deals here with stationary solutions, therefore 
the time-derivative in (\ref{eqn:dlrate}) is canceled
\begin{equation}
	\sum_j R_{ji} N_j - N_i \sum_j R_{ij} = 0.\label{eqn:crsta}
\end{equation}
Populations are normalized according to $\sum_{\text{all }j} 
N_j=1$. The algorithm solving this system or its CA counterpart is 
a classical Gauss elimination. It might not be the most efficient 
for very large systems: since the transition rates only connect 
levels from two identical or consecutive charge states $Z^\star$, 
in plasmas with a large number of possible $Z^\star$ the CR system 
(\ref{eqn:crsta}) has many zeroes and sparse matrix techniques 
may become efficient. However numerical tests (comparison of 
16- and 32-digits arithmetics, and the one proposed in Appendix 
\ref{sec:sbacch}) have shown that the accuracy of the solution is 
fair and the computation time remains reasonable even on an 
ordinary desktop computer. 

Another check will prove to be especially important in dealing 
with the configuration average validity. It is based on a 
comparison with the Saha-Boltzmann solution, and is detailed in 
Appendix \ref{sec:sbacch}. It consists in solving an 
\textit{additional rate-equation} system with the substitution of 
the rates $R^\prime$ to $R$
\begin{equation}
	R \text{\ (all processes)} \longrightarrow R^\prime
	\text{\ (collisional excitation, ionization and inverse 
	processes only).}\label{eqn:crmod}
\end{equation}

It should be noted that we do not account here for pressure-%
induced continuum lowering, and therefore the present results do 
not hold in the case of very high ion densities.

\subsubsection{Detailed balance principle and configuration average 
\label{sec:dbacav}}
When performing the accuracy check developed in Appendix 
\ref{sec:sbacch}, the ``thermodynamic solution'' --- obtained with 
partial rates $R'$ only --- turns out to be in fair agreement with 
the Saha-Boltzmann solution when one considers \textit{detailed} 
levels. An illustration is given below (Section \ref{sec:numacc}). 
However this no longer holds after \textit{configuration 
averaging}. Such a behavior may be understood from inspection of 
the microreversiblity equation
\begin{equation}
	R_{ij}^\text{p}N_i = R_{ji}^\text{q}N_j \label{eqn:micrev}
\end{equation}
where $R^\text{p}$ is the rate for some process p, $R^\text{q}$ the 
rate for the inverse process q, and $N_i$ are the \textit{level} 
populations at \textit{thermodynamical equilibrium}, thus obeying 
Saha-Boltzmann law. However, with the notations of Appendix 
\ref{sec:cfavcr}, one cannot write the ``micro''-reversibility 
equation for transitions between \textit{configurations}. Indeed, 
from the (\ref{eqn:avcrra}) rate definition, one has
\begin{align}
	R_{\alpha\beta}^\text{p}N_\alpha &= 
	 \frac{1}{\displaystyle\sum_{i\in\alpha}g_i} \left(
	 \sum_{\substack{i\in\alpha\\j\in\beta}} g_iR_{ij}^\text{p}
	 \right) \left(\sum_{i\in\alpha}N_i\right) \label{eqn:avcrde}
	 \ne \sum_{\substack{i\in\alpha\\j\in\beta}}R_{ij}^\text{p}N_i\\
	 & \ne R_{\beta\alpha}^\text{q}N_\beta\text{ in the general case}.	 
\end{align}
In the special case where \textit{all the levels of a given 
configuration would have the same transition rate}, simply 
proportional to the final level degeneracy, i.e., all transitions 
between \textit{microlevels} have identical probabilities, 
from the definition (\ref{eqn:avcrra}),
\begin{equation}
	R_{\alpha\beta}^\text{p} = g_\beta R_{ij}^\text{p} / g_j
	\text{\quad for every $i$, $j$,}\label{eqn:idrate}
\end{equation}
one can easily derive that the level populations inside a given 
configuration would simply be proportional to their degeneracy 
\begin{equation}
	N_i = g_i N_\alpha/g_\alpha,
\end{equation}
and using the rate property (\ref{eqn:idrate}) one might write the 
members of the microreversibility equation (\ref{eqn:micrev}) as
\begin{equation}
 R_{ij}^\text{p}N_i = \frac{g_j g_i}{g_\beta} 
  R_{\alpha\beta}^\text{p} \frac{N_\alpha}{g_\alpha}\text{,\quad}
 R_{ji}^\text{q}N_j = \frac{g_i g_j}{g_\alpha} 
  R_{\beta\alpha}^\text{q} \frac{N_\beta}{g_\beta}. 
\end{equation}
This clearly demonstrates that, \textit{in this circumstance}, the 
``micro''-reversibility equation $R_{\alpha\beta}^\text{p}N_\alpha=
R_{\beta\alpha}^\text{q}N_\beta$ would hold for configurations too. 
Conversely, when (\ref{eqn:idrate}) does not hold, the observed 
difference between the ``thermodynamical'' solution of
\begin{equation}
	\sum_\beta R^\prime_{\beta\alpha} N_\beta - N_\alpha 
	\sum_\beta R^\prime_{\alpha\beta} = 0 
\end{equation}
($R^\prime$ standing for collisional excitation, collisional 
ionization and their inverse processes) and the 
configuration-averaged Saha-Boltzmann solution (using average 
configuration energies) is a measure of the departure from the 
(\ref{eqn:idrate}) rule, i.e. from ``non-uniformity'' of 
transition rates inside a given configuration.

\section{Case study of CR solution in a Carbon plasma
\label{sec:carbon}}
\subsection{Atomic physics and collision processes}
A detailed CR analysis has been performed in Carbon, including 
all charge states. The list of configurations is detailed 
in Table~\ref{tab:confc}. The HULLAC computation includes 
configurations with single electron excitation up to the $n=5$ 
shell. Some configurations such as $1s^22p^4$, 
$1s^22s2p^24d$ in C\textsc{i}, $1s^22s2p4s$ in C\textsc{ii},  
or $1s^22p4p$ in C\textsc{iii} autoionize. In the moderate-$T_e$ 
range considered, it was not necessary to account for other 
multiply excited configurations, such as $1s^23lNl'$ in 
C\textsc{iii} or $2lNl'$ in C\textsc{v}. In order to consider the 
fully stripped C\textsc{vii} ion, one must include in the 
collisional-radiative equations a HULLAC fictitious configuration 
or level with zero electron population and zero energy. As seen 
in Table~\ref{tab:confc}, going from a detailed level to a 
configuration-average formalism amounts to divide the number of 
rate equations $N_\text{eq}$ by a factor of almost 12, and the  
solution of this linear-system requires a computation time 
proportional to the cube of $N_\text{eq}$.

As mentioned in Section \ref{sec:codesc}, some transition rates 
could not be computed and had to be canceled. At $T_e=10\text{ eV}$, 
8190 deexcitation rates out of 465,083 computed (1.8\%) and 178 
radiative recombination rates out of 138,195 (0.13\%) fall in this 
case. At $T_e=1\text{ eV}$, the proportion of unreliable 
deexcitation rates is somehow higher (6.2\%) while the one for 
recombination rates is unchanged. This demonstrates the (weakly) 
increasing limitation in the use of HULLAC formalism at low 
temperatures.

\subsection{Numerical accuracy of the detailed CR solution}
\label{sec:numacc}
Here as in the rest of this paper, we focus the attention on 
average net ion charges. Of course many other physical quantities 
are of interest, but if the configuration average procedure is 
correct, it should \textit{a fortiori} provide correct average 
charges. The accuracy of the solution of the modified CR system 
(\ref{eqn:crmod}) with \textit{detailed levels} is illustrated in 
Table~\ref{tab:acthsb}. Keeping in mind that the modified system 
should amount to LTE (cf. Appendix \ref{sec:sbacch}), columns 2 
and 3 of this table should be equal assuming perfect numerical 
accuracy, while column 4 should cancel. One may conclude that for 
this 1781-equation system, the matrix-inversion algorithm remains 
very efficient even with the usual 16-digit precision arithmetic 
used. Another check in 32-digit arithmetic also provides fair 
agreement with the present result.

\subsection{Validity of the configuration average}
\label{sec:valcfav}
\begin{figure}[htb]
\begin{center}
	\includegraphics[scale=0.60, angle=0]{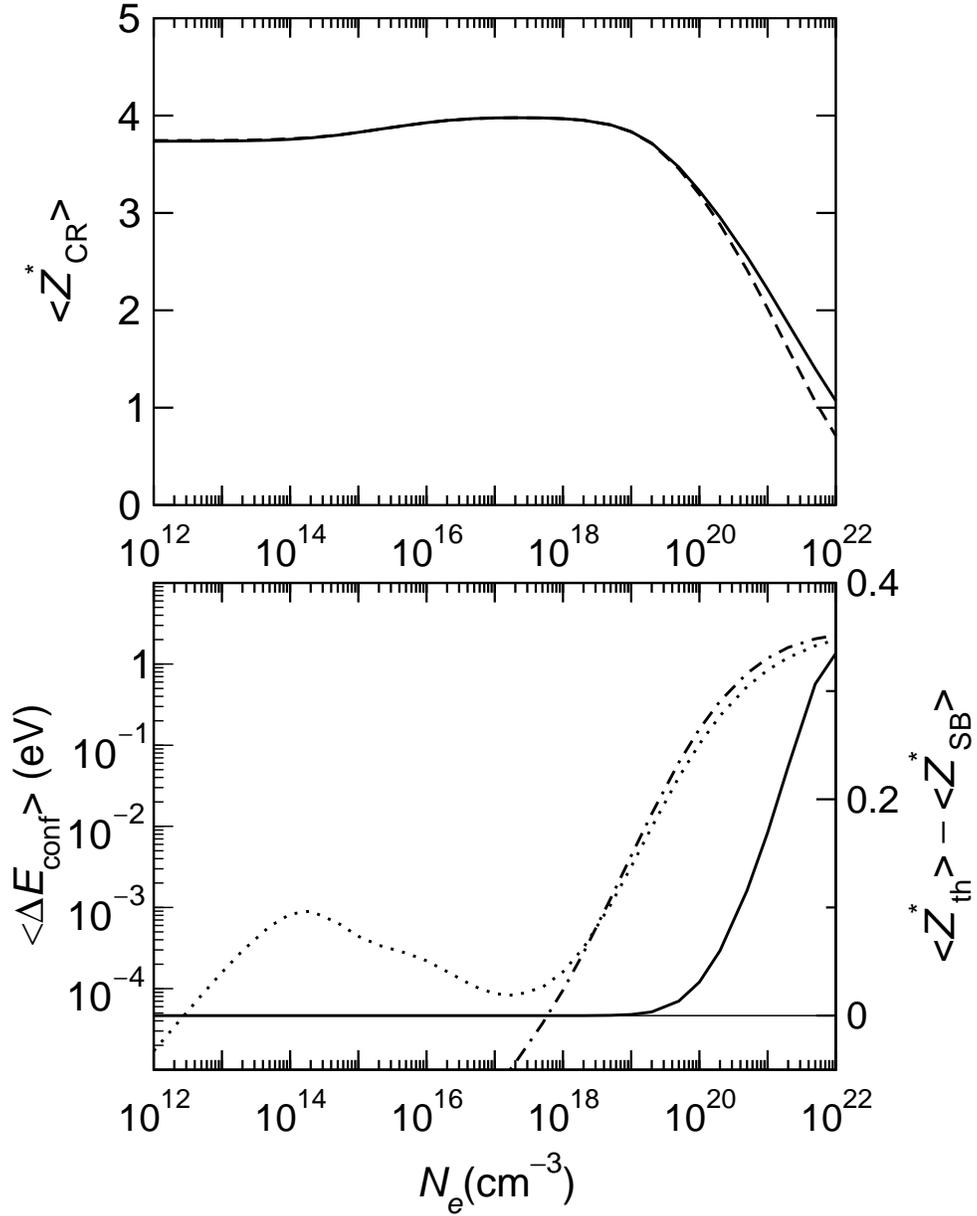}
	\caption{Upper part: Carbon plasma average charge in collisional-%
	radiative model with configuration averaging (solid line) and 
	detailed levels (broken line) as a function of electronic density 
	for an electronic temperature $T_e=10\text{ eV}$. 
	Lower part: averaged configuration energy dispersion as computed 
	from (\ref{eqn:crdise}) with collisional-radiative populations 
	(dotted line, left scale) or with Saha-Boltzmann populations 
	(dot-dash line, left scale); 
	average-charge difference (\ref{eqn:diffz}) between the modified  
	CR system $<Z^\star_\text{th}>$ and the Saha-Boltzmann solution 
	$<Z^\star_\text{SB}>$ (solid line, right scale), both computed in 
	the configuration-average formalism.\label{fig:crvcfa}}
\end{center}
\end{figure}

On the upper part of Fig.~\ref{fig:crvcfa}, we have plotted the 
average ion charge $<Z^\star>$ versus the electron density at 
$T_e=10\text{ eV}$, both in the CA and in the detailed 
collisional-radiative formalism. Therefore, both should describe 
non-LTE effects and differ from the Saha-Boltzmann value, as 
analyzed in subsection \ref{sec:dendep}. As a rule, the CA 
procedure appears as pretty accurate, which is interesting since 
this average involves a much simpler matrix inversion. But a 
detailed inspection of Fig.~\ref{fig:crvcfa} reveals some 
discrepancies, at low densities first (e.g., at 
$N_e=10^{13}\text{ cm}^{-3}$, $<Z^\star_\text{CR}>$ is 3.739 in the 
CA solution, and 3.749 in the detailed solution), and more 
prominently at densities above $N_e=10^{20}\text{ cm}^{-3}$: if 
$N_e=10^{21}\text{ cm}^{-3}$, $<Z^\star_\text{CR}>$ is 1.977 for the 
CA and 2.076 for the detailed level computation.

An intuitive explanation for this behavior is the following. In 
``coronal'' plasmas at 10 eV, the most probable charge state is the 
He-like C\textsc{v} and the most populated configuration is simply 
$1s^2$. At higher densities, recombination rates become larger and 
ions with a more complex structure dominate. Since these ions have 
configurations with a larger energy dispersion, the average 
configuration width may become of the order of magnitude of the 
thermal energy $k_BT_e$. Therefore the intuitive criterion for the 
validity of configuration averaging may be 
written as
\begin{equation}
	<\Delta E> = \sum_\alpha N_\alpha \Delta E_\alpha \ll k_BT_e
	\label{eqn:crdise}
\end{equation}
where $N_\alpha$ is the $\alpha$-configuration population (with 
$\sum_\alpha N_\alpha=1$) and $\Delta E_\alpha$ is the energy 
dispersion
\begin{equation}
	\Delta E_\alpha = \left(\frac{1}{g_\alpha}\sum_{i\in\alpha}
	g_i(E_i-E_\alpha)^2\right)^{1/2}.
\end{equation}
Unfortunately this criterion, though necessary, is far from 
sufficient. In the lower part of Fig.~\ref{fig:crvcfa}, two 
variants of the dispersion (\ref{eqn:crdise}) are plotted: the 
dotted line corresponds to populations $N_\alpha$ computed using 
the collisional-radiative solution, the dash-dot line is obtained 
with $N_\alpha$ as given by the Saha-Boltzmann equation. Both 
values are significantly below 10 eV, by a factor of 7 at the 
minimum, even for large density values where one observes a small 
inaccuracy in the configuration average approximation. Moreover we 
will exhibit a much more severe breakdown of validity of the 
criterion (\ref{eqn:crdise}) in the next section. However, in 
subsection \ref{sec:dbacav}, we have shown that a useful test 
consists in performing a collisional-radiative analysis study with 
\textit{partial} rates $R^\prime$ which should amount to the 
Saha-Boltzmann solution (i.e., LTE) if all levels in a given 
configuration had the same collision rates. Therefore, in the 
lower part of Fig.~\ref{fig:crvcfa}, the solid line is the 
difference 
\begin{equation}
	\Delta Z^\star = <Z^\star_\text{th}{}^\text{config-av}> -
	<Z^\star_\text{SB}{}^\text{config-av}>
	\label{eqn:diffz}
\end{equation}
where $<Z^\star_\text{th}{}^\text{config-av}>$ is the plasma 
charge computed with \textit{partial} rates (\ref{eqn:crmod}), 
while $<Z^\star_\text{SB}{}^\text{config-av}>$ is the plasma 
charge obtained through Saha-Boltzmann equation, \textit{both 
in a configuration-averaged formalism} (remember that such 
figures are identical when one deals with \textit{detailed} 
levels, cf.\ Table~\ref{tab:acthsb}). One notices a significant 
average-charge difference of about 0.17 at $10^{21}\text{ 
electrons/cm}^3$. As mentioned in subsection \ref{sec:dbacav}, the 
increase in the quantity (\ref{eqn:diffz}) is a clear indication 
of the breakdown of the validity of the configuration average. The 
difference (\ref{eqn:diffz}) is not a direct estimate of  
$<Z^\star_\text{CR}{}^\text{config-av}> - 
<Z^\star_\text{CR}{}^\text{detailed}>$ but a value of about 0.2 
for the test (\ref{eqn:diffz}) suggests a serious inaccuracy in 
the configuration-average approach.

\subsection{Density dependence}\label{sec:dendep}
\begin{figure}[hbt]
\begin{center}
	\includegraphics[scale=0.60, angle=0]{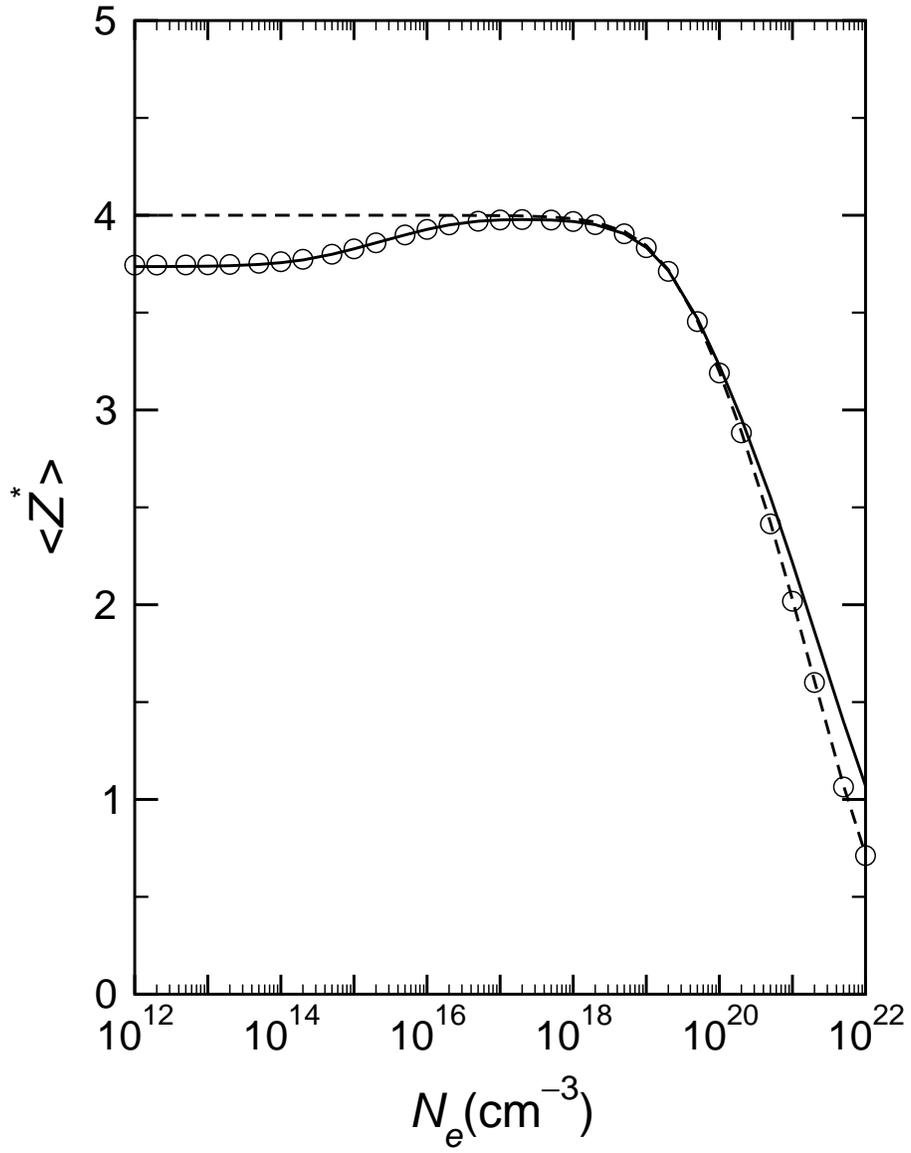}
	\caption{Carbon plasma average charge for $T_e=10\text{ eV}$. 
	Solid line: collisional-radiative solution with configuration 
	average, broken line: Saha-Boltzmann solution, circles: 
	collisional-radiative solution with detailed levels.
	\label{fig:cacrsb}}
\end{center}
\end{figure}

In order to demonstrate non-LTE effects, we have plotted on 
Fig.~\ref{fig:cacrsb} the configuration average charge of a 
10 eV-carbon plasma, both within the present collisional-radiative 
model and using the Saha-Boltzmann equation, i.e., at LTE. 
Generally speaking, as the electron density $N_e$ increases, one 
expects that LTE will prevail because collisions will dominate 
radiative processes. At high $N_e$ the most probable process 
becomes three-body recombination since it is the only one depending 
on the square of the electron density, therefore one expects a 
decrease of $<Z^\star>$ as $N_e$ increases. This is visible on 
Fig.~\ref{fig:cacrsb}, which also shows that for 
$N_e<10^{16}\text{ cm}^{-3}$, the CR charge is lower than the 
LTE charge. In coronal plasmas, dominant processes are radiative 
recombination and collisional ionization: since the latter is 
balanced by three-body recombination while the former is not 
balanced by its inverse process (if photoionization with a thermal 
radiation at $T_e$ was present, this would ensure complete LTE) 
the ``unbalanced'' recombination process tends to lower the average 
ion energy and the average plasma charge. Then the CR solution 
converges toward LTE at $N_e\simeq10^{18}\text{ cm}^{-3}$, as 
expected. A little more surprising is the divergence of the CR and 
LTE curves at higher densities $N_e>10^{21}\text{ cm}^{-3}$. Again, 
this arises from the limitation of validity of the configuration 
average, which may be observed for high densities $N_e$ as well 
as for low temperatures $T_e$ (Fig.~\ref{fig:crvcfa}). The 
\textit{detailed} CR solution (circles on Fig.\ref{fig:cacrsb}) 
agrees with the configuration-average solution at low $Ne_e$, while 
for large $N_e$ it tends to the Saha-Boltzmann limit and differs 
from the configuration average.

\section{Discussion of the configuration-average validity
\label{sec:discus}}
\subsection{Analysis of the two validity criteria}
We have demonstrated in the previous section that the energy 
criterion (\ref{eqn:crdise}) is not sufficient to ensure the 
validity of the configuration-average procedure. A physical 
explanation for this is that while the condition (\ref{eqn:crdise}) 
relies on \textit{energies} only, the analysis using the partial 
rates equation and the estimate of the difference (\ref{eqn:diffz}) 
involves the \textit{transition rates}. Therefore both conditions 
are complementary, and the latter is certainly most stringent. 
Another indication that the departure of (\ref{eqn:diffz}) from 
zero is a sign of dispersion in the transition probabilities relies 
on the analysis of the Saha-Boltzmann average charge difference 
$<Z^\star_\text{SB}{}^\text{config-av}> - 
<Z^\star_\text{SB}{}^\text{detailed}>$: as illustrated by 
Table~\ref{tab:sbcadl}, this quantity turns out to remain small 
(less than 0.1) even at low temperatures. This happens because 
the Saha-Boltzmann equation involves energies and not rates, and 
therefore is insensitive to the rate dispersion inside 
a configuration.

\subsection{Comparison with other data}
Though the aim of this work was to analyze a validity criterion 
rather than to perform a reference CR calculation in carbon, it is 
instructive to compare the present results with some recent data. 
Colgan \textit{et al} \cite{Col06} have performed a detailed 
analysis of non-LTE carbon plasmas, both in the configuration 
average and in the ``fine-structure'', i.e., detailed formalism. 
Their computation has been performed with the Cowan-based Los 
Alamos suite of codes and involves 1348 configurations and 24\ 902 
levels, much is more than in the present work. Nevertheless, as 
seen in Table~\ref{tab:Colgan}, their 10 eV-data compare 
satisfactorily with ours, whatever the density: the difference is 
less than 2\%. This is a general trend in such a range of 
temperatures, where the carbon structure tends to the very stable 
($Z^\star=4$) $1s^2$ configuration.

However, at 3 eV the difference within the detailed-formalism 
results remains reasonable (about 10\%), while large discrepancies, 
sometimes by a factor of 2, are observed on the configuration-%
average results. It is somewhat expected that low-$T_e$ results 
exhibit large dispersion, as already noticed in the NLTE 
conferences \cite{Bow03,Bow06}: then the neutral or weakly charged 
ions become preponderant, and atomic models such as HULLAC are then 
known to be much less efficient; furthermore, collision 
cross-sections are usually bigger at low $T_e$ and inaccuracies 
such as those mentioned previously (Sec. \ref{sec:codesc}) have a 
strong influence. Nevertheless the factor of 2 on the 
configuration-average plasma charge needs further attention, and is 
thoroughly analyzed in subsection \ref{sec:cascaut}.

\subsection{Low-temperature behavior}\label{sec:lowT}
An even more spectacular proof for the insufficiency of the energy 
condition is provided by the study of the carbon plasma at $T_e=1
\text{ eV}$. The results are summarized in Fig.~\ref{fig:cac1eV}.
It is noticeable that while the energy dispersion 
(\ref{eqn:crdise}), plotted as dotted or dot-dash lines, is below 
$N_e=10^{14}\text{ cm}^{-3}$ less than $T_e$, and more than 2-order 
of magnitude less at $N_e=10^{12}\text{ cm}^{-3}$, the configuration 
average validity is always dubious then, as pointed out by the 
upper part of this figure. Conversely, as stressed before, when 
the criterion on $\Delta E$ is not sufficient, the $<Z^\star>$ 
values from the modified CR system and the Saha-Boltzmann solution 
are significantly different, indicating the breakdown of the 
configuration-average procedure. It appears on this example, 
considering, e.g., $N_e=10^{12}\text{ cm}^{-3}$ and 
$10^{22}\text{ cm}^{-3}$ that both criteria should be checked.

Comparing these data to those in Table~\ref{tab:Colgan}, one 
notices that the low-density average charge in the detailed 
calculation increases with $T_e$ as expected, the $<Z^\star> 
\simeq 1$ at $T_e=1\text{ eV}$ being a plain coincidence; 
inversely, in configuration average the low-density charge turns 
out to be $<Z^\star> \simeq 2$ on a broad range of $T_e$. Of 
course the criterion (\ref{eqn:diffz}) demonstrates that this 
value is unreliable, but this unexpected ``stability'' deserves a 
more detailed analysis.

\subsection{Breakdown of the configuration-average validity 
for cascade autoionization processes}\label{sec:cascaut}
\begin{figure}[htb]
\begin{center}
	\includegraphics[scale=0.60, angle=0]{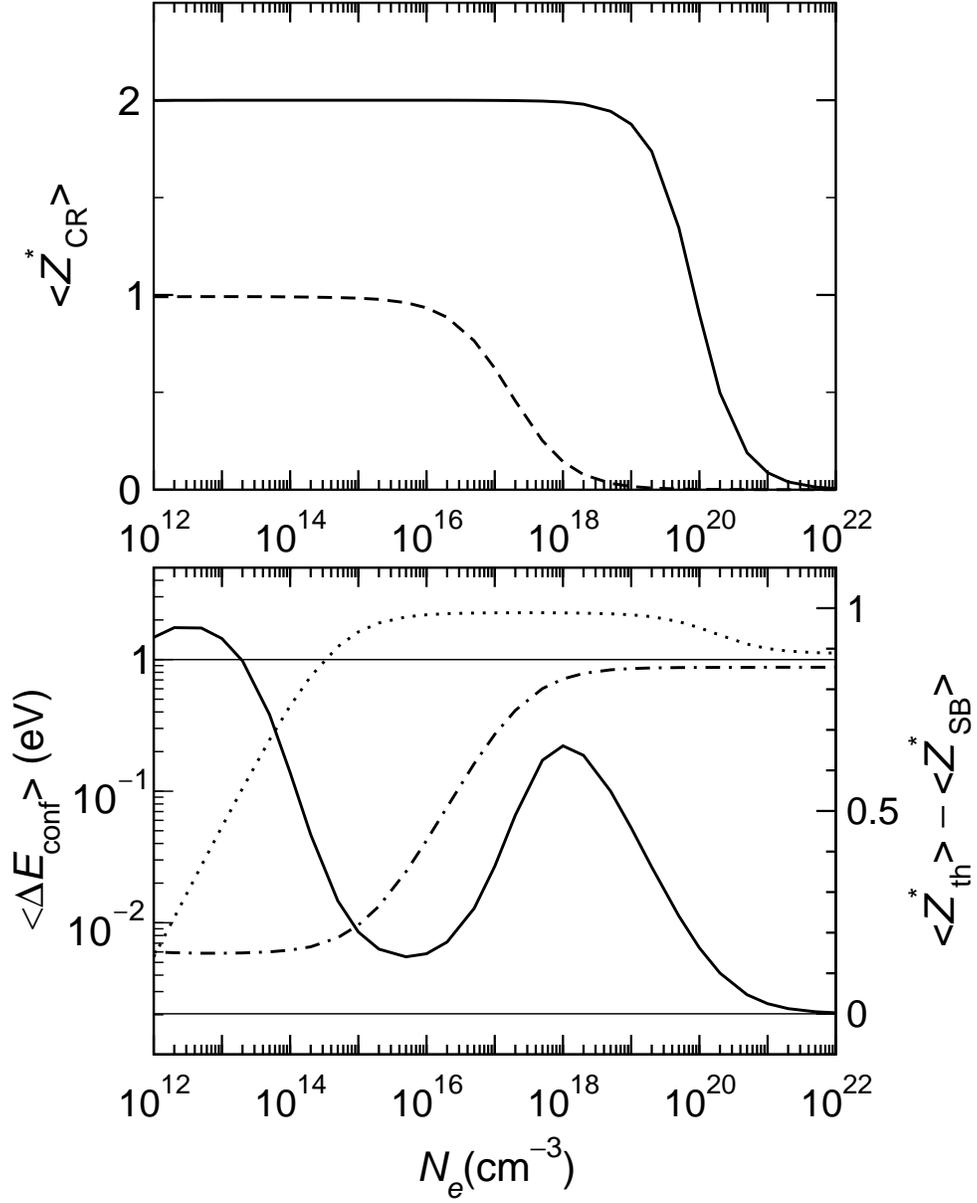}
	\caption{Same as figure \ref{fig:crvcfa}, but for an electron 
	temperature $T_e = \text{1 eV}$. Notice on the top subfigure the 
	strong departure from the configuration-average validity if 
	$N_e \le 10^{14}\text{ cm}^{-3}$, which is not detected by the 
	$<\Delta E_\text{conf}>$	criterion (bottom subfigure, dotted 
	line), but is detected by the criterion on $<Z^\star_\text{th}> 
	- <Z^\star_\text{SB}>$ (bottom subfigure, solid line).
	\label{fig:cac1eV}}
\end{center}
\end{figure}

\begin{figure}[htb]
\begin{center}
\ifx\pdfoutput\undefined
	\includegraphics[scale=0.55, angle=-90]{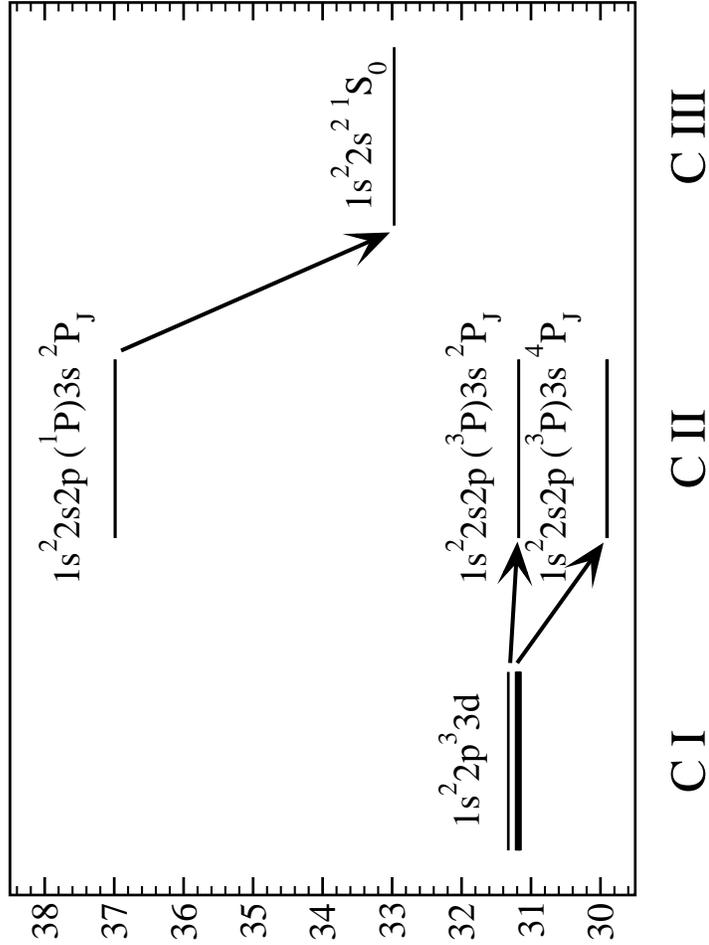}
\else
	\includegraphics[scale=0.55, angle=0]{CI-III_autoion}
\fi
\caption{
  Autoionization processes involving three particular 
  configurations in C\textsc{i}, C\textsc{ii}, C\textsc{iii}. The 
  arrows indicate the dominant transitions accounted for in the 
  present calculation. The vertical axis is the energy in eV, 
  respective to the C\textsc{i} ground state. Only some of the 38 
  levels of the $1s^2 2p^3 3d$ configuration are displayed. About 
  ten of them autoionize toward the $1s^2 2s 2p\;({}^3P) 3s\ {}^2P$ 
  or the $1s^2 2s 2p\;({}^3P) 3s\ {}^4P$ levels --- which are below 
  the C\textsc{ii} limit $1s^22s^2\;{}^1S_0$ and do not themselves 
  autoionize  ---, and none toward the upper levels ($1s^2 2s 
  2p\;({}^1P) 3s\ {}^2P$), which are the only two levels in the 
  configuration $1s^2 2s 2p 3s$ that autoionize. No cascade 
  autoionizing process exists from $1s^2 2p^3 3d$ to $1s^2 2p^2$, 
  though large \textit{average} autoionization rates connect these 
  configurations. 
  \label{fig:cascauto}}
\end{center}
\end{figure}

A thorough examination of the various-configuration influence has 
revealed that the unexpected low-$T_e$ low-$N_e$ configuration-%
averaged value $<Z^\star> \simeq 2$ arises mostly from \textit{two} 
configurations, namely $1s^2 2p^3 3d$ in C\textsc{i} and 
$1s^2 2s 2p 3s$ in C\textsc{ii}. As a matter of fact, a simple 
model involving the five configurations $1s^2 2s^2 2p^2$, 
$1s^2 2p^3 3d$, $1s^2 2s^2 2p$, $1s^2 2s 2p 3s$, and $1s^2 2s^2$ 
fairly reproduces the unexpected behavior of Fig.~\ref{fig:cac1eV}. 
In a rather unusual way, the dominant transition rates are then 
the \textit{autoionization} rates from some levels of the first 
configuration to some levels of the second one. Dielectronic  
recombination rates are accounted for here, but they are small 
because the electron density is low. As illustrated by 
Fig.~\ref{fig:cascauto}, while the highest members of the 
$1s^2 2p^3 3d$ configuration decay to the 
$1s^2 2s 2p ({}^3P) 3s\,{}^2P$ doublet or to the 
$1s^2 2s 2p ({}^3P) 3s\,{}^4P$ quartet, all these levels lie 
below the C\textsc{ii} first ionization limit and do not 
autoionize; conversely, the \textit{upper} member of this 
configuration, namely the $1s^2 2s 2p ({}^1P) 3s\,{}^2P$ doublet, 
does autoionize towards the C\textsc{iii} ground state 
$1s^2 2s^2\,{}^1S$ with a very large probability. So this 
detailed-level analysis shows that the \textit{cascade 
autoionization process} $1s^2 2p^3 3d \rightarrow 1s^2 2s 2p 3s 
\rightarrow 1s^2 2s^2$ is \textit{not} allowed. However, the 
\textit{configuration-average} autoionization rates $1s^2 2p^3 3d 
\rightarrow 1s^2 2s 2p 3s$ and $1s^2 2s 2p 3s \rightarrow 
1s^2 2s^2$ are both very large ($6.8\times 10^{14}\text{ s}^{-1}$ 
and $3.6\times 10^{13}\text{ s}^{-1}$ respectively) and 
introduce a spurious very intense \textit{two-step transition} 
from C\textsc{i} to C\textsc{iii}, which explains the very stable 
and large C\textsc{iii} ground state population. The explanation 
why Colgan \textit{et al} \cite{Col06} did not notice such a big 
discrepancy as ours between the detailed and CA ionization stages 
at 3 eV appears now clearly: the $1s^2 2p^3 3d$ configuration was 
not accounted for in their calculation. Remarkably enough, 
adding or removing dozens of configurations other than the ones 
listed in Fig.~\ref{fig:cascauto} does not significantly change 
the average $Z^\star$ (CA or detailed), while a plain 
five-configuration model qualitatively reproduces the 
Fig.~\ref{fig:cac1eV} behavior. 

Though the $1s^2 2p^3 3d$ and $1s^2 2s 2p 3s$ configurations have 
an energy dispersion $\Delta E_\alpha$ larger than $T_e$ (3.41 eV 
and 2.93 eV respectively) these configurations have very small 
populations ($9.7\times10^{-16}$ and $8.8\times10^{-11}$ 
respectively), therefore the criterion (\ref{eqn:crdise}) is 
largely fulfilled. A more strict criterion such as 
$\max_\alpha(\Delta E_\alpha) < k_B T_e$ could be proposed, but it 
would be hardly satisfied except at very large $T_e$ 
($\max(\Delta E_\alpha) = 5.3\text{ eV}$ for the carbon 
configurations considered here), and furthermore one may question 
whether configurations with very small populations should 
contribute to this condition. 

To conclude this low-$T_e$ analysis, it must be noted that the 
current computation generates levels from $1s^2 2p^3 3d$ and 
$1s^2 2s 2p 3s$ configurations at very close energies (explaining 
the large autoionization rate as is usual for this process), 
which may not be reproduced by more accurate computations or other 
atomic models. Concerning experimental data, in the NIST (National 
Institute of Standards and Technology) tables \cite{NISwww}, one 
finds the $1s^2 2s2p ({}^3P) 3s$ levels (but not the 
$2s2p({}^1P) 3s$ ones), while the highly excited configuration 
$1s^2 2p^3 3d$ is absent; the agreement on the 
$1s^2 2s2p ({}^3P) 3s$ level positions is good. Noticeably, most 
of the autoionizing states considered here are absent in the NIST 
tables, probably because their broadening makes their position 
difficult to report. Nevertheless, the present analysis is useful 
in itself because the reported effect must happen in other 
configurations or ions as soon as these quasi-degeneracies are 
indeed present.

\section{Conclusion}\label{sec:concl}
This work initially stemmed from the need of a collisional-radiative 
solver (or postprocessor) for the HULLAC code; at the time of 
writing it, such solver was not part of the HULLAC suite, at least 
in the publicly distributed code. Few CR computations based on the 
HULLAC suite have appeared recently \cite{Chu06}. The possibility 
of performing the configuration average makes the handling of the 
CR equations considerably easier, the system size decreasing from 
1781 to 149 in the analyzed carbon case. Superconfiguration codes 
\cite{Han06} provide an interesting alternative concerning the 
computation efficiency and their ability to deal with complex 
atoms. However the inclusion in these formalisms of effects such as 
a full configuration interaction is still lacking. Of course 
the detailed-level methods will for long suffer from the 
considerable computing time, considering that the carbon plasma 
case analyzed here is one of the simplest. In this work we have 
proposed a method to check the validity of the configuration 
average procedure which goes far beyond the plain criterion on 
energy dispersion. It involves the solution of two 
\textit{configuration-averaged} CR systems instead of one, but 
this is much less cumbersome than solving a detailed-level CR 
system. This criterion is based on rate equations for collisional 
excitation and ionization and their reverse processes, and is 
therefore sensitive to a possible non-uniformity inside a 
configuration of these rates only. However, it may be generalized 
to include other processes, for instance the radiative processes, 
provided that one accounts for absorption and spontaneous emission 
from a fictitious Planck radiation field in thermal equibrium with 
the electrons. In the analyzed case of cascade autoionization 
processes, it has been demonstrated that a spectacular breakdown 
of the configuration-average validity may occur which can only be 
detected using the criterion proposed here. Finally, the present 
formalism should also be efficient in computing other physical 
quantities such as the opacity, the emissivity \cite{Col06} or 
radiative cooling coefficients \cite{Chu06} of non-LTE plasmas; 
such a topic has been recently addressed in our laboratory 
\cite{dGD06}.

\begin{acknowledgments}
The authors gratefully acknowledge stimulating discussions with 
T. Blenski and the irreplaceable assistance of M. Busquet in 
the usage of the HULLAC code. They also thank A. Bar-Shalom, 
M. Klapisch, and J. Oreg for making this code available.
\end{acknowledgments}

\appendix
\section{The averaged collisional-radiative equations
\label{sec:cfavcr}}
Rate equations for detailed levels $i$ belonging to a given 
configuration $\alpha$ are written as (\ref{eqn:dlrate})
while their \textit{configuration average} is defined as
\begin{equation}
	\frac{dN_{\alpha}}{dt} = \sum_{\beta} R_{\beta\alpha} N_{\beta} 
	 - N_{\alpha} \sum_{\beta} R_{\alpha\beta}
\end{equation}
where the configuration population is $N_{\alpha} = 
\sum_{i\in\alpha} N_i$ and the transition rates from 
configuration $\alpha$ to $\beta$ writes
\begin{equation}
	R_{\alpha\beta} = \frac{1}{g_{\alpha}}\sum_{i\in\alpha} 
	  \sum_{j\in\beta} g_i R_{ij}\label{eqn:avcrra}
\end{equation}
$g_{\alpha}$ being the degeneracy $\sum_{i\in\alpha} g_i$ 
of the initial configuration.

\section{Checking the numerical CR system solution 
\label{sec:sbacch}}
The CR analysis in the detailed-level case (\ref{eqn:crsta}) or 
its configuration-average counterpart requires the solution of  
a potentially very large system of linear equations. One should 
therefore check the numerical accuracy of this procedure. In this 
CR system, the $R_{ij}$ stand for all the transitions processes 
enumerated in section \ref{sec:HULLAC}. Because no external 
electromagnetic field is present, two of these processes are not 
balanced by their inverse processes: radiative deexcitation and 
radiative recombination. In order to check the accuracy of these 
CR equations, one substitutes \textit{partial rates} 
$R^\prime_{ij}$ to the full rates $R_{ij}$, which only include 
\textit{collisional excitation, collisional ionization and their 
inverse processes}. 

In the \textit{detailed-level} case, since the effects accounted 
for by rates $R^\prime$ obey the detailed-balance principle, 
the correct numerical solution of this system --- numerically as 
complex as the original system since the same number of equations 
is involved --- must be the Saha-Boltzmann solution. 

In the \textit{configuration-average} case, the comparison of this 
solution to the Saha-Boltzmann solution with averaged energies 
provides a validity check of the CR solution as discussed, e.g., in 
subsection \ref{sec:valcfav}.

\clearpage
\bibliography{colrad}





\clearpage
\section*{Tables}
\enlargethispage{1.5cm}
\begin{table}[ht]
\begin{center}
\caption{List of configurations included in the HULLAC-based 
collisional-radiative analysis of a carbon plasma. $[1s^2]$ means 
unrepeated identical core. One has $3\le N\le5, 0\le l\le N-1$ 
for each charge state. $N_\text{conf}$ is the number of 
configurations, $N_\text{lev}$ the number of (relativistic) 
levels.\label{tab:confc}}
\bigskip
\begin{tabular}{@{\extracolsep{5pt}} c l c c}
\hline\hline
Charge & \multicolumn{1}{c}{Configurations} & $N_\text{conf}$ & 
$N_\text{lev}$ \\
\hline
C\textsc{i} & $[1s^2]\ 2s^22p^2,\ 2s2p^3,\ 2p^4,\ 2s^22pNl,
\ 2s2p^2Nl,\ 2p^3Nl$ & 39 & 1004 \\
C\textsc{ii} & $[1s^2]\ 2s^22p,\ 2s2p^2,\ 2p^3,\ 2s^2Nl,
\ 2s2pNl,\ 2p^2Nl$ & 39 & 513 \\
C\textsc{iii} & $[1s^2]\ 2s^2,\ 2s2p,\ 2p^2,\ 2sNl,\ 2pNl$ & 27 & 
166 \\
C\textsc{iv} & $[1s^2]\ 2s,\ 2p,\ Nl$ & 14 & 24 \\
C\textsc{v} & $1s^2,\ 1s2s,\ 1s2p,\ 1sNl$ & 15 & 49 \\
C\textsc{vi} & $1s,\ 2s,\ 2p,\ Nl$ & 15 & 25 \\
C\textsc{vii} & & 1 & 1\\
\hline
Total && 150 & 1782 \\
\hline\hline
\end{tabular}
\end{center}
\end{table}

\begin{table}[bh]
\begin{center}
\caption{Accuracy check of the collisional-radiative (CR) 
\textit{detailed-level} solution for carbon at $T_e=10\text{ eV}$. 
The average $<Z^\star_\text{th}>$ is the ion charge obtained by 
solving the modified CR system (\ref{eqn:crmod}), and should be 
equal to $<Z^\star_\text{SB}>$, the average ion charge derived 
from Saha-Boltzmann equation, for infinite numerical accuracy. 
The additional test $\delta_\text{max}$ is the maximum difference 
on the ion-level populations between this modified CR system and 
the Saha-Boltzmann solution.\label{tab:acthsb}}
\bigskip
\begin{tabular}{@{\extracolsep{10pt}} c c c c}
\hline\hline
$N_e\text{(cm$^{-3}$)}$ & $<Z^\star_\text{th}>$ & 
$<Z^\star_\text{SB}>$ & $\delta_\text{max}$ \\
\hline
   $10^{12}$ &  4.000004022 & 4.000004022  & $7.7\times10^{-14}$ \\
   $10^{14}$ &  3.999998278 & 3.999998278  & $3.2\times10^{-14}$ \\
   $10^{16}$ &  3.999823805 & 3.999823805  & $3.3\times10^{-12}$ \\
   $10^{18}$ &  3.982612437 & 3.982612437  & $3.2\times10^{-10}$ \\
   $10^{20}$ &  3.192620512 & 3.192620504  & $4.8\times10^{-9}$ \\
   $10^{22}$ &  0.711283477 & 0.711283469  & $3.8\times10^{-10}$ \\
\hline\hline
\end{tabular}
\end{center}
\end{table}
\clearpage

\enlargethispage{3cm}
\begin{table}[thb]
\begin{center}
\caption{Average carbon plasma charge from Saha-Boltzmann equation 
at $T_e=10\text{ eV}$: configuration-average and detailed-level  
values.\label{tab:sbcadl}}
\vspace{15pt}
\begin{tabular}{@{\extracolsep{5pt}} c c c }
\hline\hline
$N_e\text{(cm$^{-3}$)}$ & $<Z^\star_\text{SB}{}^\text{config-av}>$ 
& $<Z^\star_\text{SB}{}^\text{detailed}>$ \\
\hline
   $10^{12}$  &  4.000004  &  4.000004 \\
   $10^{13}$  &  4.000000  &  4.000000 \\
   $10^{14}$  &  3.999998  &  3.999998 \\
   $10^{15}$  &  3.999982  &  3.999982 \\
   $10^{16}$  &  3.999824  &  3.999824 \\
   $10^{17}$  &  3.998240  &  3.998240 \\
   $10^{18}$  &  3.982613  &  3.982612 \\
   $10^{19}$  &  3.844529  &  3.844448 \\
   $10^{20}$  &  3.194906  &  3.192621 \\
   $10^{21}$  &  2.026326  &  2.017892 \\
   $10^{22}$  &  0.717721  &  0.711283 \\
\hline\hline
\end{tabular}
\end{center}
\end{table}

\begin{table}[hbt]
\caption{Average ionization of a carbon plasma: Colgan \textit{et 
al} \cite{Col06} and present work. The Colgan's ``fine-structure'' 
(FS) is a detailed formalism, CA is the configuration average in 
both works.}\label{tab:Colgan}
\begin{tabular}{cc@{\qquad}c@{\ \ }c@{\qquad}c@{\ \ }c}
\hline\hline
$T_e \text{(eV)}$ & $N_e \text{(cm$^{-3}$)}$ & 
 \multicolumn{2}{c@{\qquad}}{Colgan \textit{et al}} &
 \multicolumn{2}{c}{This work} \\
 && FS & CA & detailed & CA\\
 \hline
 3 & $10^{13}$ & 1.730 & 1.486 & 1.902 & 1.999 \\
   & $10^{15}$ & 1.923 & 1.895 & 2.037 & 2.166 \\
   & $10^{17}$ & 1.952 & 1.948 & 1.991 & 2.501 \\
   & $10^{19}$ & 1.004 & 0.959 & 1.179 & 1.980 \\\hline
10 & $10^{13}$ & 3.723 & 3.701 & 3.746 & 3.739 \\
   & $10^{15}$ & 3.862 & 3.856 & 3.828 & 3.828 \\
   & $10^{17}$ & 3.979 & 3.978 & 3.976 & 3.976 \\
   & $10^{19}$ & 3.786 & 3.785 & 3.833 & 3.835 \\
\hline\hline
\end{tabular}
\end{table}

\clearpage
\enlargethispage{3cm}
\section*{Figures}

\end{document}